\documentclass[aps,floatfix,prl,superscriptaddress,reprint,showpacs,10pt,preprintnumbers,longbibliography]{revtex4-1}
\usepackage[utf8]{inputenc}
\usepackage[pdftex]{graphicx}
\usepackage{float}
\usepackage{amsmath}
\usepackage{amssymb}
\usepackage{amsfonts}
\usepackage{dsfont}
\usepackage{array}
\usepackage{bm}
\usepackage{mathrsfs}
\usepackage{pifont}
\usepackage{multirow}
\usepackage{upgreek}
\usepackage{xcolor}
\usepackage{standalone}
\usepackage[pdftex,
            pdftitle={Tackling critical slowing down using global correction steps
  with equivariant flows: the case of the Schwinger model},
            pdfauthor={Jacob Finkenrath},
            bookmarks,
            colorlinks,
            linkcolor=myblue,
            citecolor=mymagenta,
            menucolor=black,
            urlcolor=myblue,
            plainpages=false,
            pdfpagelabels,
            hypertexnames=false]{hyperref}
\usepackage{verbatim}
\usepackage{slashed}
\definecolor{mymagenta}{RGB}{200, 0, 100}
\definecolor{myblue}{RGB}{45, 48, 146}

\graphicspath{{figures/}}

\begin{document}
\title{Tackling critical slowing down using global correction steps
  with equivariant flows: the case of the Schwinger model}
\author{Jacob Finkenrath}  
\affiliation{Computation-based Science
  and Technology Research Center, The Cyprus Institute, 20
  Konstantinou Kavafi Str., 2121 Nicosia, Cyprus}

\date{\today}
\begin{abstract}
We propose a new method for simulating lattice gauge theories in
the presence of fermions. The method combines flow-based generative
models for local gauge field updates and hierarchical updates of
the factorized fermion determinant. The flow-based generative
models are restricted to proposing updates to gauge-fields within
subdomains, thus keeping training times moderate while increasing
the global volume. We apply our method performs to
the 2-dimensional (2D) Schwinger model with $N_f=2$ Wilson Dirac fermions and show that no critical slowing down is observed in the sampling of
topological sectors up to $\beta=8.45$. Furthermore, we show that
fluctuations can be suppressed exponentially with the distance
between active subdomains, allowing us to achieve acceptance
rates of up to $99\%$ for the outer-most accept/reject step on
lattices volumes of up to $V=128\times128$.
\end{abstract}

\maketitle

\textit{Introduction. --}  Gauge field theories are solved  non-perturbatively 
 by  defining them  on a discrete spacetime lattice and carrying out a numerical evaluation,
provided their infinite volume and zero lattice spacing limits are
taken. These limits require simulation at fine lattice spacing, in
order to access logarithmic corrections to discretization
effects~\cite{Husung:2021mfl}. For example, in Quantum Chromodynamics (QCD),
accessing quantities that can connect to experiments in the precision
frontier requires sub-percent accuracy after combining statistical and systematic
errors. Estimating reliably systematic errors includes  extrapolation to the continuum limit. This systematic error is reduced by having  lattice
spacings smaller than 0.05~fm that is typically  the smallest lattices spacing accessible to  current simulations.

The Hybrid Monte Carlo (HMC) algorithm~\cite{Duane:1987de,Gottlieb:1987mq} has played an essential role in allowing for large scale simulations of lattice gauge
theories. However, a major drawback is the so-called critical slowing down observed in simulations as 
the lattice spacing decreases. Critical slowing down is referred to as 
 the exponential increase of the computational cost
required for HMC to explore topological sectors of the theory as the
lattice spacing is reduced~\cite{Schaefer:2010hu}, thus resulting in a
Markov Chain with long trajectories effectively that remain (frozen)  in the same
topological sector.

In Ref.~\cite{Luscher:2009eq}, it was shown how constructing a map
that can trivialize a gauge-theory can be employed to solve
topological freezing. A recent development~\cite{Kanwar:2020xzo} was
the application of  a flow-based generative model built of machine
trainable affine coupling layers that can be used to construct such a map
showing how topological freezing is overcome as criticality is
approached. A major obstacle  in such machine learning approaches,
however, is their scalability with the  volume. Namely, while
the number of degrees of freedom of the theory increases linearly with
the volume, the trainable parameters and therefore the training time and
memory requirements of the coupling layers scale polynomially. Indeed,
as the volume is increased and more trainable parameters are
introduced, convergence of the optimizer becomes non-trivial and
requires extensive fine-tuning or pre-training
techniques~\cite{DelDebbio:2021qwf}.

\begin{figure}
  \includegraphics[width=1.\linewidth]{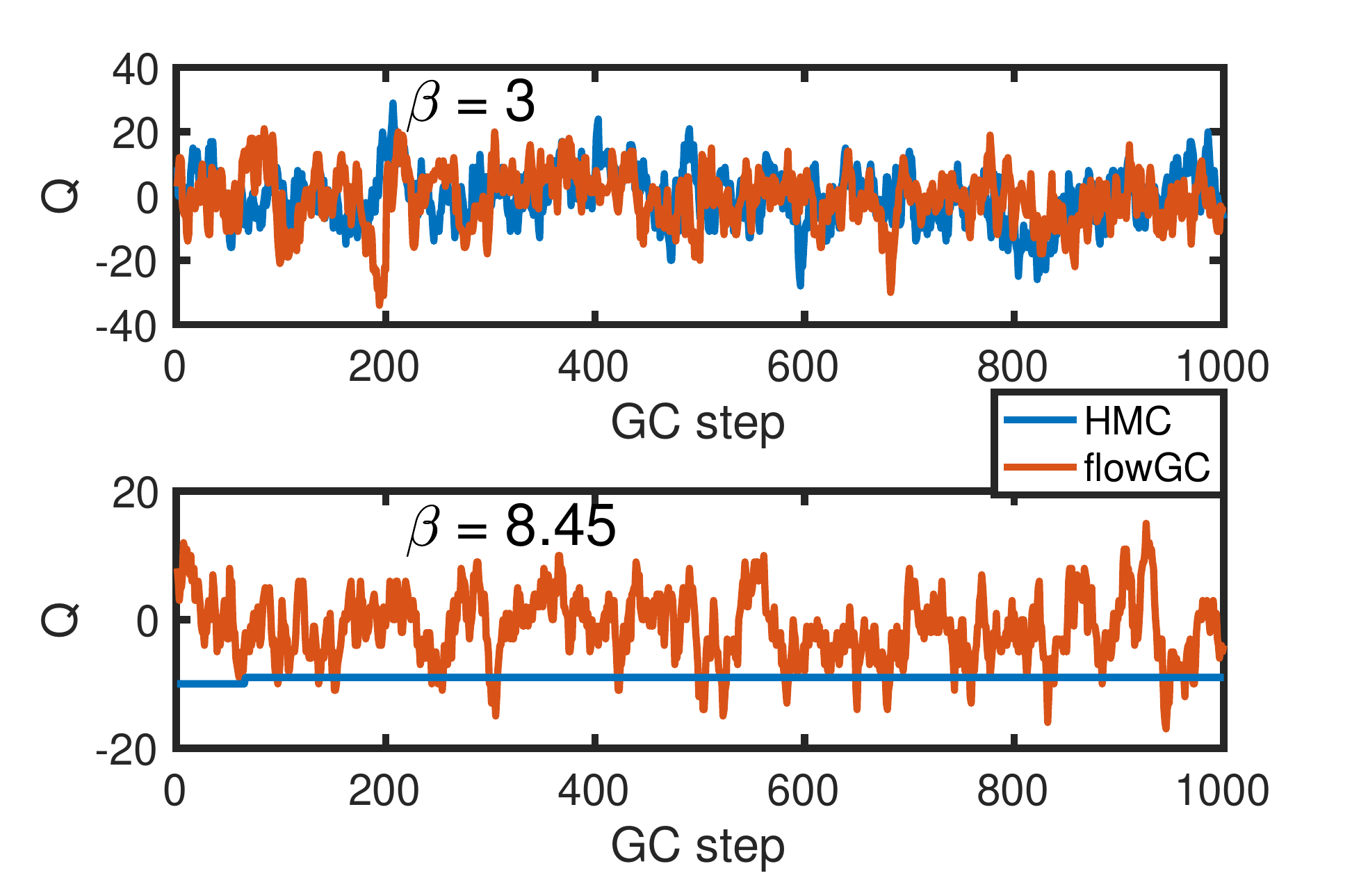}
  \caption{Monte Carlo history of the topological charge at a coarse
    lattice with $\beta=3$ (upper panel) and a fine lattice with
    $\beta=8.45$ (lower panel) at a volume of $V=128\times 128$
    generated using the HMC algorithm and the flowGC method proposed.}
  \label{fig:topo}
\end{figure}

In this work, we  introduce a novel approach that exploits locality so that the flow-based model
proposes updates to subdomains of the lattice. Such approaches have
traditionally been  key drivers in the simulation of lattice gauge theories
and lattice QCD in particular, allowing simulations at physical quark
masses.  Recent examples are  multi-grid approaches
\cite{Frommer:2013fsa,Luscher:2004,Luscher:2007se}, efficient
computation of the trace of the inverse Dirac operator using
hierarchical probing~\cite{Stathopoulos:2013aci}, and  multi-level
algorithms~\cite{Ce:2016ajy,Ce:2016idq}, 
The  new approach proposed here  combines flow-based
generative models for proposing gauge-filed configurations~\cite{Kanwar:2020xzo} and
hierarchical updating of the gauge-fields for including the fermion
determinant~\cite{Finkenrath:2012az}. We illustrate that applicability of   this method in the case of the  2D
Schwinger model with two flavors of degenerate Wilson fermions
($N_f=2$). We show that this approach  mitigates critical slowing down
but also keeps high acceptance rates for
volumes as large as 128$\times$128. In Fig.~\ref{fig:topo}, we
demonstrate that  our method, denoted as ``flowGC'',
crucially increases the rate at which topological sectors of the
theory are explored at very fine lattice spacings ($\beta=8.45$) where
HMC is effectively frozen.

\textit{Schwinger model --} The Boltzmann factor of the discrete 2D
Schwinger model can be written as
\begin{equation}
  \rho(U) = \frac{1}{Z} \textrm{exp}\{- \beta S_g(U) + 2 \cdot \textrm{ln} \, \textrm{det} D(U) \}~.
  \label{eq:pathintegral}
\end{equation}
For the ultra local pure gauge action, we employ the plaquette action
$\beta S_g = \beta \sum_{i=1}^V \textrm{Re} [P(x)]$, where $V=L\cdot L$ is the lattice volume with lattice extent $L$ in each direction and
$\beta$ is the gauge coupling. The plaquette at point $x$ is given by
$P(x) = U^\dagger_2(x)
U^\dagger_1(x+\hat{2})U_2(x+\hat{1})U^\dagger_1(x)$ with gauge links
$U_\nu(x) \in U(1)$, which connects the point $x+\hat{\nu}$ with
$x$. The plaquette is gauge invariant under an arbitrary gauge
transformation $U_\nu(x) \rightarrow g(x+\hat{\nu}) U_\nu(x) g(x)$
with $g(x) \in U(1)$.  The topological charge takes integer values in
the Schwinger model and can be defined by $Q=1/(2\pi) \sum_{i=1}^V(
\arg(P(x) )$, where $\arg(e^{i \theta})=\theta$.  The fermion action is
given by the determinant of the Wilson Dirac operator $D(U) \in
\mathbb{C}^{2V\times 2V}$, which can be represented as a complex
sparse matrix with a real determinant, see
e.g. Ref.~\cite{Christian:2005yp}.

\textit{Corrections via Metropolis accept/reject steps --}
Given a process $T_0(U\to U')$, which allows proposing a new sample
$U'$ starting from a previous sample $U$ with known distribution
$\tilde{\rho}(U')$ and for which detailed balance is satisfied, a set
of configurations $\{U\}$ weighted with a fixed point distribution
$\rho(U)$ can be generated via a Markov chain. This is given by a combination
of a proposal with a subsequent Metropolis accept/reject step  
\begin{equation}
\begin{split}
 0) &\quad \text{Propose $U'$ according to $T_0(U\to U')$}\\
 1) &\quad P_{acc}(U\to U') =  \text{min} \left[1, {\frac{\tilde{\rho}(U) \rho(U')}{\rho(U) \tilde{\rho}(U')}} \, \right].
\end{split}
\label{eq:prop_correct}
\end{equation}
Most Markov Chain Monte Carlo algorithms used for large scale simulations  of
lattice gauge theories, such as the HMC algorithm, are
based on this approach, where the Metropolis accept/reject step works
as a global correction (GC).  In general, the Boltzmann factor $\rho$
depends on extensive quantities, i.e.~the actions are extensive
quantities, such as the corresponding variances scale with the
physical volume.  Assuming that the ratio of distributions
$(\tilde{\rho}(U) \rho(U'))/(\rho(U) \tilde{\rho}(U'))$ is log-normal
distributed, then for  the acceptance rate $P_{acc}$ of
Eq.~\eqref{eq:prop_correct}, it  follows that
\begin{equation} P_{acc} = \textrm{erfc}\{\sqrt{\sigma^2(\Delta S)/8}\}
\end{equation}
with the variance $\sigma^2(\Delta S)$, where $\Delta S= \textrm{ln}\{
\rho(U')\} - \textrm{ln} \{ \rho(U) \} +
\textrm{ln}\{\tilde{\rho}(U)\} - \textrm{ln} \{\tilde{\rho}(U') \}$~\cite{Knechtli:2003yt}.
The extensive character of $\Delta S$ implies
that the acceptance rate drops exponentially as the volume increases.
 In order to achieve high acceptance rates, it is necessary to
minimize the variance $\sigma^2(\Delta S)$, which can be done
\begin{itemize}
    \item [1.] by using correlations between $\rho$ and $\tilde{\rho}$
    \item [2.] by reduction of the degrees of freedom of $\rho$ and $\tilde{\rho}$.
\end{itemize}
For   HMC,  case 1 above applies, namely the distance of $\rho$ and
$\tilde{\rho}$ of the distribution of the Shadow Hamiltonian 
\cite{Clark:2007ffa,Kennedy:2012gk}, is minimized. A combination of
both cases leads to a generalization of the GC step of
Eq.~\eqref{eq:prop_correct}, which is achieved by introducing a
hierarchy of filter steps. If we factorize the target fixed point
distribution $\rho(U)$ into $n+1$ parts $P_i$ with
\begin{equation}\label{eq:factor}
  \rho_{j}(U)=P_0(U,\{\alpha^{(0)}\})\,P_1(U,\{\alpha^{(1)}\})\,\dots P_{j}(U,\{\alpha^{(j)}\})\,,
\end{equation}
where $j\le n$, $\rho_n(U)\equiv \rho(U)$ and $\{\alpha^{(j)}\}$ are
arbitrary sets of parameters, then the GC step in
Eq.~\eqref{eq:prop_correct} splits into $n$ successive steps, with the
$j$\textsuperscript{th} given by
\begin{align}
  P^{j}_{acc}(U\to U') = & \text{min} \left[1, \frac{\rho_{j-1}(U) \rho_{j}(U')}{\rho_{j}(U) \rho_{j-1}(U')} \right]\nonumber\\
  = & \text{min}\left[1, \frac{P_{j}(U',\{\alpha^{(j)}\})}{P_{j}(U,\{\alpha^{(j)}\})} \right].
\end{align}
Now we can introduce a hierarchy of nested accept/reject steps
which can be iterated to filter out local fluctuations effectively.

\textit{Flow-based generative models and trivializing maps --} We
employ gauge equivariant maps ($m(U)$) as in
Ref.~\cite{Kanwar:2020xzo}, constructed by combining coupling layers
$m=\prod_j g_j$. The map is required to transform gauge-fields from a
trivialized phase of our gauge model with distribution
$\rho_{trivial}$ into a non-trivial distribution $\tilde{\rho}$,
i.e. $m(U):\,\rho_{trival}(U) \to \tilde{\rho}(U)$. Each coupling
layer $g_j$ transforms a set of active gauge links using a set of
gauge-invariant objects, such as plaquettes. The maps can be
represented by tunable convolutional networks with few hidden layers
\cite{Albergo:2019eim,dinh2017density,rezende2016variational}. Keeping
track of the phase space deformation, this transformation can be
computationally simplified using active, passive, and static masks,
such that the Jacobian of the transformation becomes triangular. This
yields a tractable determinant of the Jacobian of the transformation,
which is needed to train the parameters of the coupling layers.

Now, we can write the distribution of the generative model as
\cite{Albergo:2019eim,Kanwar:2020xzo}
\begin{equation}
    \tilde{\rho}(U) = \rho_{trival}(m^{-1}(U)) \prod_{j} \textrm{det} J(g_j^{-1}(\{\alpha_{j}^{(0)}\}))
\end{equation}
with the trivial distribution $\rho_{trivial}$ and the Jacobian
$J(g_j^{-1}(\{\alpha_{j}^{(0)}\}))$ of each coupling layer. The
parameters $\{\alpha_{j}^{(0)}\}$ are tunable weights in the coupling
layers.

Minimizing directly the variance of the accept/reject step in
\eqref{eq:prop_correct} requires pairs of configuration $U,\,U'$, with
$U$ distributed via $\rho(U)$, which are \textit{apriory} not
available.  It turns out that minimizing the difference between
$\tilde{\rho}$ and ${\rho}$ is sufficient and leads to the
definition of the loss-function as the Kullback-Leibler divergence
\begin{equation}
  \textrm{loss}(U) = \textrm{ln} (\tilde{\rho}(U)) - \textrm{ln} ({\rho(U)}),
  \label{eq:loss}
\end{equation}
which can be now minimized by training through iteratively drawing
random samples with $\rho_{trivial}$ and adjusting the weights
$\alpha_{i,j}^{(0)}$ in the coupling layers.
 
\begin{figure}
\includegraphics[width=1.\linewidth]{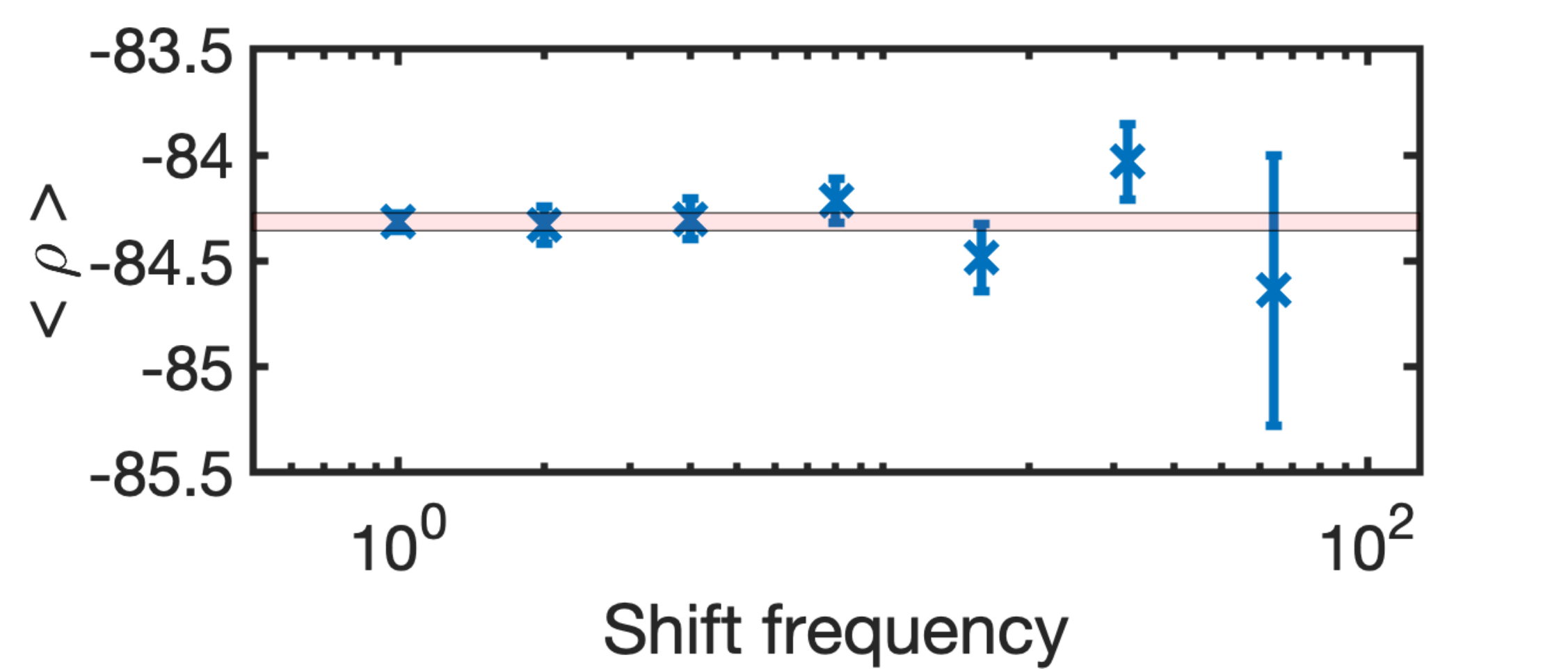}
      \caption{The flow distribution $\tilde{\rho}$ as a function of
        the number of iterations between random shifts of the
        lattice. For comparison the result for 1 shift per iteration
        is drawn with the horizontal band.}
      \label{fig:frq}
\end{figure}

\textit{Domain decomposition of gauge equivariant flows in the 2D
  Schwinger model. --} In Refs.~\cite{Kanwar:2020xzo,Albergo:2021vyo},
equivariant flows were introduced for generating field configurations
of the 2D pure-gauge Schwinger model, demonstrating improved sampling
of topological sectors at large values of the coupling, namely
$\beta$=7, where HMC fails. Suppressing volume fluctuations within
$\sigma^2(\textrm{loss})$, however, proved challenging for lattices with
$L>16$, which is the interesting case. We  will show that this can be addressed by splitting the
accept/reject step in Eq.~(\eqref{eq:prop_correct}) using domain
decomposition and, thus, limiting the dimension of $\tilde{\rho}$ to the
size of the domains. Because the pure gauge action is ultra local,
updates of links within domains can be carried out independently of
other domains if links that lie in or at the boundary of the domains
are kept constant. We employ equivariant flows that are trained to
generate link variables within the domain, given a fixed set of links
connecting the domains.
To ensure ergoticity, we periodically shift the lattice by a random
translations $T_{\vec{x}}: \vec{x_0} \to \vec{x_0}+ \vec{x}$, similar
to Ref.~\cite{Luscher:2004pav}. While the lattice action $\langle
\rho(U) \rangle = \langle \rho (T_{\vec{x}}(U)) \rangle$ is invariant
under translation, the trained gauge equivariant map is \textit{apriori} not $\langle\tilde{\rho}(U) \rangle \neq
\langle\tilde{\rho}(T_{\vec{x}}(U))\rangle$. This means that after
each shift, to calculate $\tilde{\rho}(U)$ we need to apply the
reverse map $m^{-1}:U \to U_{trivial}$ to obtain the prior
distribution. As an empirical check, we measure
$\langle\tilde{\rho}(U) \rangle$ as we vary the frequency by which we
shift the lattice, as shown in Fig.~\ref{fig:frq}. 
We see that
translations result in small fluctuations of the resulting
distribution, indicating that no violation in translational invariance
with statistical significance is  observed.

To train the flow with fixed boundary conditions, we start from the
software and workflow that implements the periodic boundary conditions
followed in Ref.~\cite{Albergo:2021vyo}. Namely, we train a flow
using periodic boundary conditions for a lattice with dimensions
$l\times l$, with $l$ being the lattice extent of the domains,
i.e. the global lattice is constructed from $n\times n$ domains with
$n=L/l$. The flow is then re-trained allowing only the links which do
not enter the plaquettes and span boundaries to be updated.
While training, the configurations in each batch use
different boundary links but are kept constant between training
iterations or epochs of one era. For each era, we re-generate boundary
links using a flow trained with periodic boundary conditions. We use a
batch size of 4096 and 1000 epochs per era. Indicatively, for
$\beta=8.45$, training took about 14 hours on an NVIDIA V100 GPU,
reaching an acceptance rate of 28\% using $l=8$. With such a flow,
trained for local updates and to which we can add random translations
 with minor effect on the resulting density, it is straightforward
to generate ensembles of larger lattices, as demonstrated in
Fig.~\ref{fig:topo} for $L$=128. Note that in order to change
topological sectors, there is a lower bound on the physical domain
size, which for this case ($l/\sqrt{\beta} = 8/\sqrt{8.45}$) is
satisfied.

\textit{Global correction steps using fermions. --} Integration of the
fermions in the path integral yields the determinant of the Wilson
Dirac operator as in Eq.~\ref{eq:pathintegral}, a non-local
operator. Nonetheless, the fermion action can be splitted in a way
that allows using Eq.~\eqref{eq:factor} via a recursive Schur
decomposition,
\begin{equation}
    \textrm{det} D(U) = \textrm{det}\, \mathcal{S}(U) \cdot \prod_i  \textrm{det} \,D(U^{(1)}_i) 
    \label{eq:Schur}
\end{equation}
with the Schur complement that is defined on the even blocks,
$\mathcal{S}(U) = 1-D^{-1}(U^{(1)}_{e,e}) D(U^{(1)}_{e,o})
D^{-1}(U^{(1)}_{o,o}) D(U^{(1)}_{o,e})$ and the block Dirac operators
$D(U^{(1)}_i)$ defined on a block. The superscript $(1)$ denotes the
first level decomposition since the Schur decomposition can be applied
recursively to multiple levels. Here we  only use a single
level. The decomposition effectively factorizes the long range modes,
which are captured by the Schur complement, from the short range
modes, captured by the block operators. Note that as in the
gauge-field domain decomposition, the block operators only depend on
links within a domain, that means they  decouple exactly from each other.  The
procedure is perfectly suited to be used in a hierarchy of
accept/reject filtering steps because it introduces additionally a
computational cost ordering. The accept/reject steps of the block
operators can be done in parallel and can be iterated to filter out
larger local fluctuations of the determinant. In this work, the
problem sizes are sufficiently small for LU-decomposition to be used
for the determinant calculation. For systems with more degrees of
freedom however, such as in lattice QCD, the determinant ratio can be
estimated stochastic, as discussed in Ref.~\cite{Finkenrath:2012az}.

The Schur complement $\mathcal{S}(U)$ contains the interactions
between the domains and, therefore, scales with the volume as an
extensive quantity. The acceptance rate of the global correction step, thus,
decreases exponentially with the volume. We mitigate this by
introducing parameters to exploit correlations between the different
factors of the action. Namely, a shift in the gauge coupling can be introduced between the pure-gauge action and the
determinant 
\begin{equation}
   S(\beta,\Delta \beta) =  (\beta + \Delta \beta) S_g + S_f
\end{equation}
with $S_f=-2 \cdot \textrm{ln} \, \textrm{det} D(U)$.  The variance of
the action is minimized when $\Delta \beta$ fulfills the relation
\begin{equation}
  \Delta \beta = - \textrm{cov}(S_g,S_f)/\sigma^2(S_f),
\end{equation}
with $\textrm{cov}(x,y)$ the covariance of $x$ and $y$. One way to use
this feature is, for example, to generate flow based updates of the
domains using $\beta + \Delta \beta$ as the coupling and reweighting
back to the target coupling $\beta$ during the accept/reject step of
the global correction. In our algorithm, which as will be
explained includes four accept/reject steps, we generalize this
approach by allowing for a different $\Delta \beta$ at each step~\cite{Finkenrath:2012az}. Details of the steps
and the values of $\Delta \beta$ used are given in the Supplemental
Material.

\begin{figure}
  \includegraphics[width=1.\linewidth]{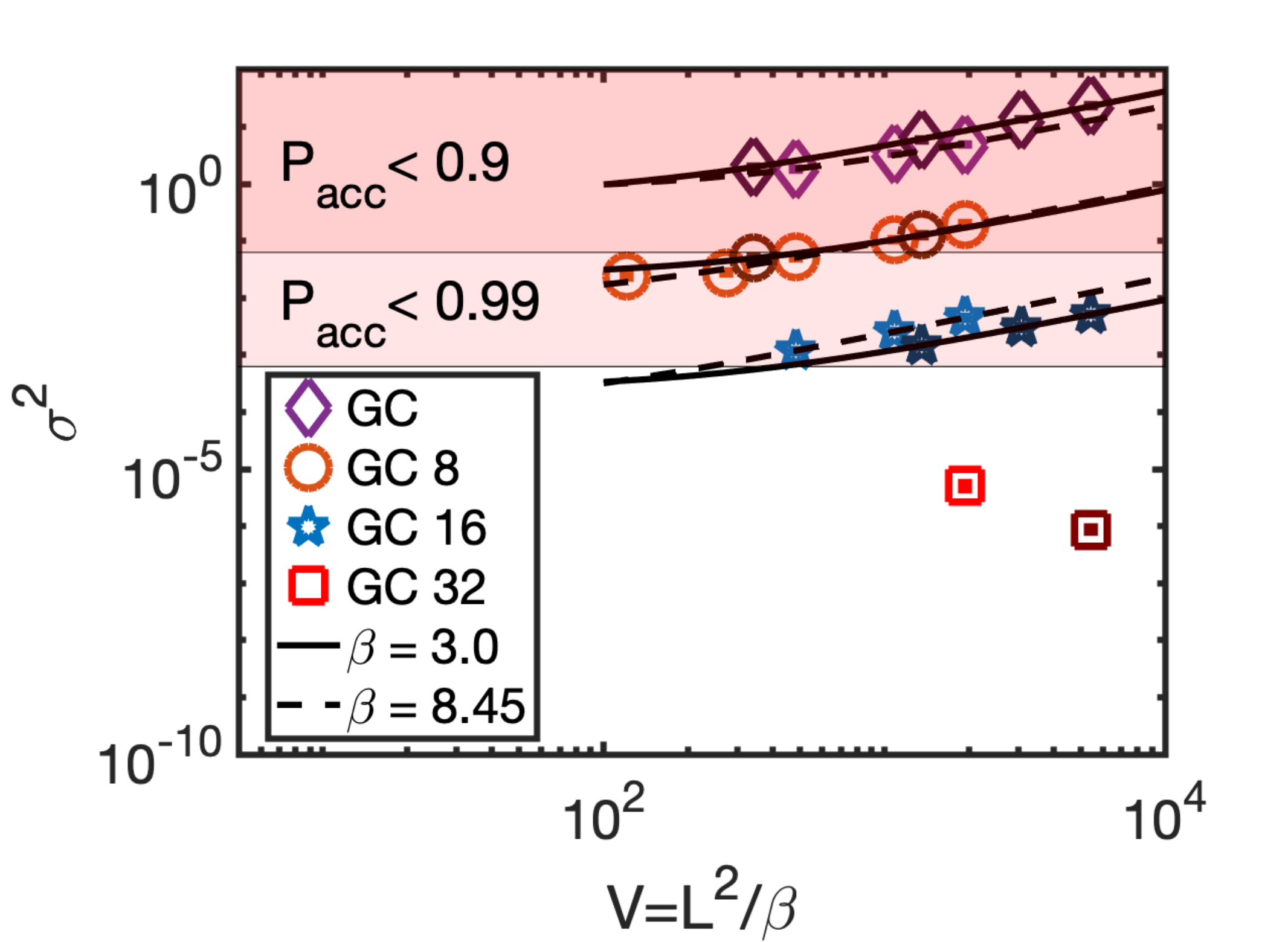}
  \caption{The variance of the global correction step as a function of
    the global volume ($V/\beta$) for $\beta=3$ (darker points) and
    $\beta=8.45$ (lighter points). We show the naive global correction
    step \textit{GC} without hierarchical filtering of the block and
    extended block operator (diamonds), as well as for the 5-level
    flowGC algorithm with block distance $L_{b}=8$ (circles), with
    $L_{b}=16$ (stars), and with $L_{b}=32$ (square).}
  \label{fig:acceptance}
\end{figure}

To further improve the global acceptance rate we note that domains
decouple effectively exponentially with their distance via $ \propto
\textrm{exp}(-c_0 \; m_{PS} \cdot |x-y|)$ and the effective decoupling
length, therefore, depends  on the lowest physical mode, i.e. the
pseudoscalar mass $m_{PS}$~\cite{Luscher:2003vf,Ce:2016ajy}. We, thus, increase the distance between domains being updated by
generalizing the checkerboard coloring of the blocks to four colors
and only update domains of same color, while all others are kept
constant. In two dimensions this is possible if the decomposition is
such that it yields an even number of domains in each direction. The
global acceptance rate can now be further increased by introducing an
additional intermediate step, which includes corrections from the
eight blocks surrounding the block being updated, i.e. a $3L_b\times
3L_b$ operator, with the active $L_b\times L_b$ block located in the
center.  We model the variance of the global GC step as
\begin{equation}
\sigma^2 (V,\beta, d, m_{PS}) = \frac{A \cdot V}{m_{PS} \beta^{3/2}}
\cdot \textrm{exp}\{ - B\cdot d \; m_{PS} \},
\end{equation}
where $d$ is the distance between active domains and find $A=0.0030(1)$ and
$B=2.4062(59)$. From now on, the method described will be referred to
as a 5-level flowGC algorithm, which includes the following steps:
\begin{itemize}
    \item [0.] Flow proposal to generate 100 samples within each
      active block with $l=8$
    \item [1.] Accept/reject step over the 100 samples using the pure
      gauge action of the active blocks as target probability and
      keeping the final accepted configuration. The acceptance rate is
      $P^{(1)}_{acc}\sim 0.25$.
    \item [2.] Calculation of the determinant of the block operator
      $D(U_j)$ with $L_b=n\cdot l$ with $n \in 1,2,\ldots$ and accept/reject. Repeat, starting from step 0. and repeat two to four
      times. The acceptance rate with $n=2$ is $P_{acc}^{(2)} \sim 0.7$.
    \item [3.] Calculation of the extended $3L_b \times 3L_b$
      Dirac operator and performance of an accept/reject step. Repeat, starting from step 0. twice. 
    The acceptance rate is $P_{acc}^{(3)} \sim 0.75$.
    \item [4.] Calculation of Schur complement term performing a global
      accept/reject step correcting to the target probability $\propto
      \textrm{exp}\{- \beta S_g+S_f \}$.
\end{itemize}
The significance of including steps 2. and 3. should be stressed
here, since not including them results in an acceptance rate that quickly
decreases with increasing volume, as shown in Fig.~\ref{fig:acceptance}. Note that for the 5-level flowGC algorithm
studied here, 15.6\% of the gauge links are updated in each global
iteration step. When varying the block sizes of the fermion
determinant domains, $L_b$, the total ratio does not change since we
keep the flow based proposal fixed to $l=8$ blocks for the flow update
in step 1. Therefore, the improvement of the acceptance rate observed as
we increase the domain block sizes ($L_b=8$, $L_b=16$, and $L_b=32$ in
Fig.~\ref{fig:acceptance}) is due to the increased distance between
active domains and the effectiveness of the filtering applied in step 3. From this
analysis, we conclude that for a lattice of size $L=128$ we can achieve
$P_{acc}>0.97$ for distances $d\gtrsim 16$.

\begin{table}[]
    \caption{The coupling $\beta$ and bare fermion mass $m_0$ used for
      three representative $N_f=2$ ensembles of the total
      generated. We also show the measured plaquette ($P$) and
      topological charge ($Q^2$). For the ensembles with gauge
      couplings $\beta \in [1,\, 6]$ we set the parameters as in
      Ref.~\cite{Christian:2005yp}, while for the finer lattice
      $\beta=8.45$ we refer to Ref.~\cite{Albandea:2021lvl}.}
    \label{tab:paras}
    \centering
    \begin{tabular}{c|c|c|c|c}
        $\beta$ &  $m_0$ & $a m_{PS}$   &  $P$ & $Q^2$ \\
          3.0   & -0.082626  &   0.2241(38)       &  0.82705(12)      & 67.6(112)   \\
          6.0   & -0.034249  &   0.1649(23)       &  0.91659(10)      & 24.3($\phantom{0}$24)  \\
          8.45  &  $\phantom{-}$0.0$\phantom{00000}$  &   0.1951(17)       &  0.94084($\phantom{0}$7)    & 21.9($\phantom{0}$25)   \\
    \end{tabular}
\end{table}

\begin{figure}
  \includegraphics[width=1.\linewidth]{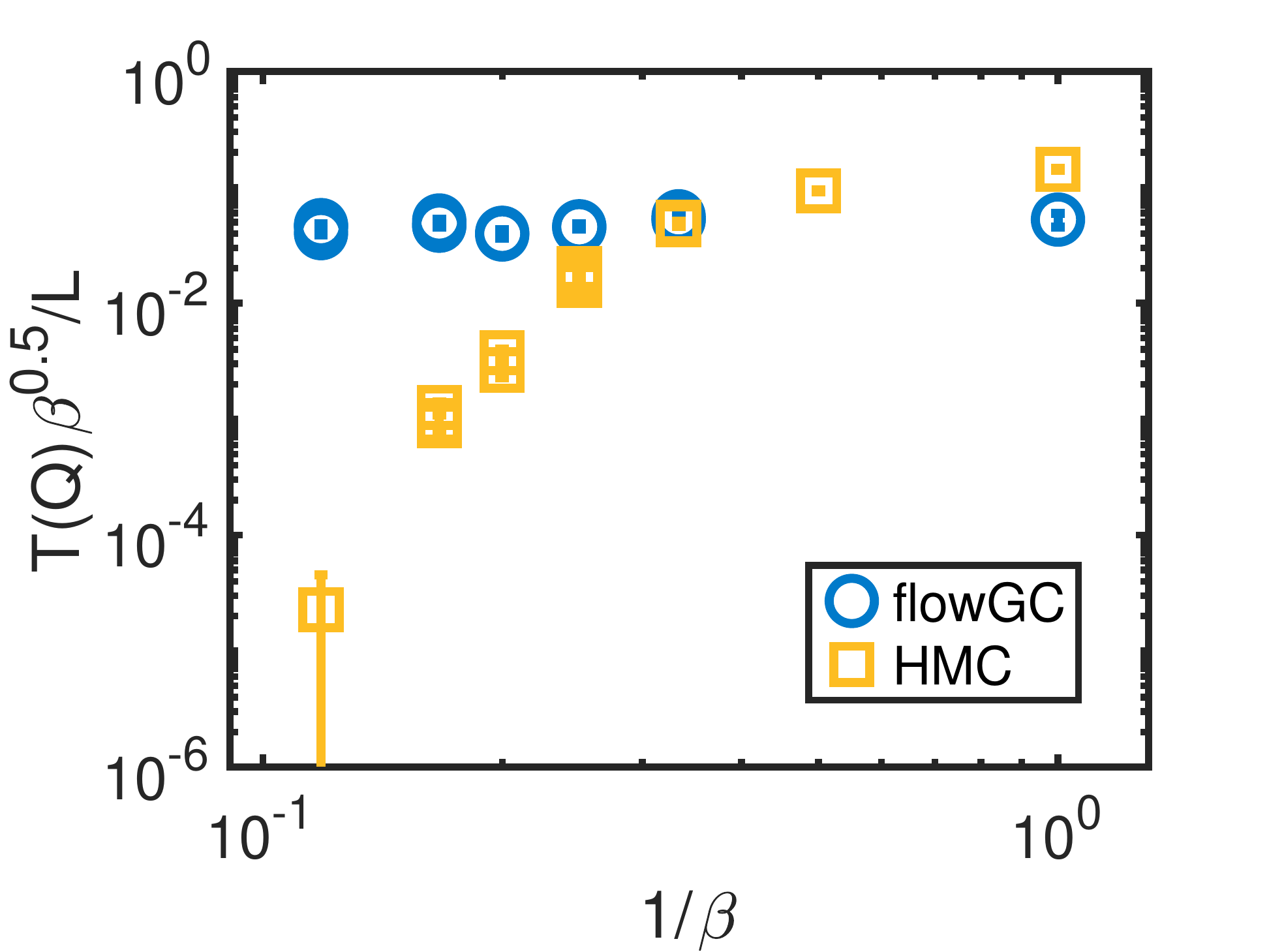}
      \caption{Tunneling rate of the topological charge per update versus
        the inverse coupling for the 5-level flowGC algorithm
        (circles) with $l=8$ flow proposals and $L_b=16$ and the HMC
        algorithm (squares). We normalize the tunneling rate with the
        volume and scale with $\sqrt{\beta}$ to derive a dimensionless
        quantity.}
      \label{fig:histo}
  \end{figure}

\textit{Markov Chain Monte Carlo simulations with dynamical
  fermions. --} We generate Markov chains with $\beta \in [1.0, \,
  8.45]$ and lattice sizes of $L/\beta \in [22,\, 74]$. The gauge
coupling range spans from values where topological sampling is
possible using HMC ($\beta \in [1,\, 4]$) to values where with HMC we
observe the onset of critical slowing down ($\beta \in [4, \, 6]$) as
well as values of $\beta > 6.00$, where the HMC algorithm effectively
freezes. In Table~\ref{tab:paras}, we list the parameters and
observables measured for three values of $\beta$ that are representative of
these three regions. To compare how well the topological charge is
sampled, we measured the so-called tunneling rate per global
acceptance step via
\begin{equation}
    T(Q) = \langle | Q_i - Q_{i+1} | \rangle.
\end{equation}
We choose this quantity rather than the auto-correlation time to
compare the topological sampling behavior of the different algorithms,
since it increases with the lattice extent $L/\sqrt{\beta}$. It  is
also easier to estimate for cases with few changes in the topological
charge, namely for HMC when $\beta=8.45$. For the HMC, $Q_i$ are
separated by one molecular dynamics update with the acceptance rate
tuned to $P_{acc}>90\%$, while for the 5-level flowGC we separate with
one global correction step and use a gauge flow with $l=8$ domains,
meaning at most 15.6\% of the links are updated per step. As shown in
Fig.~\ref{fig:histo}, the tunneling rate for the 5-level flowGC
algorithm remains constant over the $\beta$ values simulated, and is
consistent with the rate achieved by HMC at $\beta=3.0$. HMC appears
to be more favorable for the cases $\beta<2$, however only by a factor
of at most two, while the 5-level flowGC achieves over two orders of
magnitude better tunneling rate for $\beta>6$.

\textit{Conclusion. --} We  present a novel algorithm for effectively
simulating lattice gauge theories with fermions. The algorithm comprises of  gauge
equivariant flows for updating sub-domains of the global lattice
combined with hierarchical blocking of the fermion determinant
followed by a global correction step. We derive a 5-level flowGC
algorithm, which is shown to  overcome critical slowing down in the
Schwinger model with two degenerate flavors of fermions. In addition, domain
decomposition allows scaling to large lattice sizes illustrating the effectiveness of the algorithm by 
 simulating up to $128\times 128$ lattices. The variance of
the global correction step is found to decrease exponentially with the
distance $d$ between active blocks, i.e.~$\sigma^2 \propto
V/(m_{PS}\beta^{3/2}) \cdot \textrm{exp}(- 2 d \cdot m_{PS})$, with
$m_{PS}$ the pseudoscalar mass. Such a dependence with $d$ enables us to tune the block
sizes $L_b$ for a given minimum distance $d$. We demonstrate the
approach by comparing the tunneling rate of the topological charge
achieved by our method and HMC and show that the 5-level flowGC
algorithms achieves orders of magnitude higher tunneling rate at
$\beta=8.45$ where HMC freezes.

The 5-level flowCG  algorithm can be extended to  4D and gauge theories of larger gauge
groups, such as  lattice QCD. However, a number of additional  challenges may need to be addressed, such as   a possible
deterioration of the acceptance rate due to the larger block
sizes~\cite{Finkenrath:2012az} that may be needed for decoupling
domains in simulations with physical quark masses.
Including the block determinant into the
training, as in Ref.~\cite{Albergo:2021bna} could address this issue.
Furthermore, the  algorithm can be made  more efficient when
combined with HMC, similar what is discussed in
Ref.~\cite{Albandea:2021lvl}. It may be possible, if topological
freezing is resolved, to restore the sampling rate as expected from
the Langevin class of algorithms, which might increases with the
inverse lattice spacing squared
$a^{-2}$~\cite{Baulieu:1999wz,Luscher:2011qa}.

\begin{acknowledgments}
\textit{Acknowledgments.}  
The author acknowledges Giannis Koutsou for careful reading the manuscript and detailed  discussions.  
J.F.~received financial support by the
PRACE Sixth Implementation Phase (PRACE-6IP) program (grant agreement
No.~823767) and by the EuroHPC-JU project EuroCC (grant agreement
No.~951740) of the European Commission.  Parts of the runs were
performed on the Cyclone machine hosted at the HPC National Competence Center of
Cyprus at the Cyprus Institute.
\end{acknowledgments}

\bibliography{bibliography}

\newpage
\appendix
\begin{widetext}
\begin{center}
\textbf{\large Supplemental Material: Details on runs with global fermionic correction steps}
\end{center}
\end{widetext}

\begin{figure*}
  \includegraphics[width=0.84\linewidth]{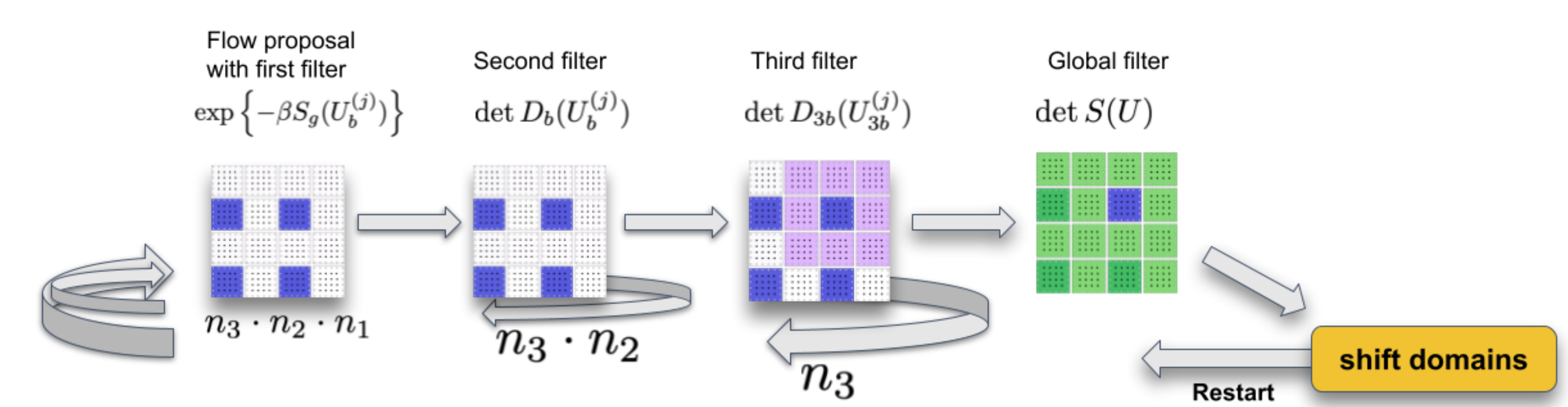}
      \caption{The schema illustrates the different levels  of the 5 level flowGC algorithm.}
      \label{fig:schema}
\end{figure*}

\textit{Details on runs with 5 level flowGC algorithm. -- }
For the global correction step algorithm driven by the gauge equivariant flow the target Boltzmann weight is given by
\begin{equation}
    P(U) = Z^{-1} \left( \prod_j^{N_f} \textrm{det} D_j(U) \right) e^{-\beta S_g(U)}
\end{equation}
with the partition sum $Z$, which we will drop from now on.
Now, using domain decomposition we can split the action into
\begin{equation}
\begin{split}
    \rho(U) =& \quad P^{(4)} \cdot \prod_{j=1}^{N_b}  \left( P^{(3)}_{j}  P^{(2)}_{j} \prod_{k=1}^{N_l} P^{(1)}_{k,j} \right) \\
    =& \quad \textrm{det} \, D^2(U) \cdot \textrm{exp}\{ -\beta S_g(U) \}
\end{split}
\end{equation}
where the parts $P^{(3)}_{j},P^{(2)}_{j},P^{(1)}_{k,j}$ does not include links of other active
domains $i$ with $i \neq j$.

Now, we can introduce a 5 level hierarchical filter step with the distributions
\begin{widetext}
\begin{equation}
\begin{split}
 0) &\quad \tilde{\rho} = \rho_{trival}(m^{-1}(U_l^{(k,j)})) \prod_{i=0}^{N_{coupling}} \textrm{det} J(g_i^{-1}(\{\alpha_{i}^{(0)}\}))  \\
 1) &\quad P^{(1)}_{k,j} = \textrm{exp}\left\{ - \left(\beta - \sum_{i=2}^{N_{lvl}} \delta \beta_{i}^{(1)}\right) S_g(U_l^{(k,j)}) \right\} \\
 2) &\quad P^{(2)}_{j} = \textrm{exp} \left\{ 2\left(1-\sum_{i=3}^{N_{lvl}} \delta \beta_{i}^{(2)}\right) \cdot \textrm{ln} \, \textrm{det} \, D_b(U_{L_b}^{(j)}) - \delta \beta_2^{(1)}  \sum_{k=1}^{N_{l}}  S_g(U_{l}^{(k,j)})\right\} \\
 3)  &\quad P^{(3)}_{j} = \textrm{exp} \left\{ - \delta \beta_{4}^{(3)} \cdot \textrm{ln} \, \textrm{det} \, D_{3b}(U_{3L_b}^{(j)}) + \delta \beta_{3}^{(2)} \cdot \textrm{ln} \, \textrm{det} \, D_b(U_{L_b}^{(j)}) - \delta \beta_3^{(1)} \sum_{k=1}^{N_{l}} S_g(U_{L_b}^{(k,j)})\right\} \\
 4)  &\quad P^{(4)} = \textrm{exp} \left\{ 2 \cdot \textrm{ln} \, \textrm{det} \, S(U) +  \sum_{j=1}^{N_b} \left[ \delta \beta_{4}^{(3)} \textrm{ln} \, \textrm{det} \, D_{3b}(U_{3L_b}^{(j)}) + \delta \beta_{4}^{(2)} \textrm{ln} \, \textrm{det} \, D_b(U_{L_b}^{(j)}) - \delta \beta_4^{(1)} \sum_{k=1}^{N_{l}} S_g(U_{L_b}^{(k,j)})\right]\right\} 
\end{split}
\label{eq:distr_lvl}
\end{equation}
\end{widetext}
with $N_{lvl}$ the number of levels, $N_l$ the number of subdomains within a block of length $L_b$ and $N_b$ the number of active blocks.
 Step 0) to 3) can be performed for each block $j$ or $k$ independently while only step 4) includes global correlations.
Note that if $L_b>8$ the domain $L_b\times L_b$ is further decomposed into $l \times l$ subdomains with $l=8$. Step 0) and 1)
is then performed on each subdomain $k$ individual. 
Each step can be iterated as illustrated in Fig.~\ref{fig:schema}.

\textit{Parameter tuning. --}
Now, the parameters $\delta \beta_i^{(j)}$ can be tuned by minimizing the variance of the higher level accept/reject steps
utilizing the co-variance between the different action parts, see also  Ref.~\cite{Finkenrath:2012az}.
We can write the action of the $i$th level step as
\begin{equation}
 S_i(U) = \sum_{j=0}^{i} \beta_i^{(j)} S^{(j)}(U), \qquad \quad i=1,2,\ldots,n
\end{equation}
with the difference of the actions
\begin{equation}
 \Delta_i = S_i(U')-S_i(U).
\end{equation}
Now, we can introduce a cost-ordered hierarchy, where the more expensive, larger term, such as the
Schur complements do not enter the low level accept--reject steps.
This implies $\beta_i^{(j)} = 0$ for $i<j$. Moreover, the additional parameters
have to sum up to the target parameters, namely $\sum_{i=1}^n \beta_i^{(j)} = \beta^{(j)}$.

Now, minimizing the variance starting from the top level leads to a coupled linear systems
which can be exactly solved. The system of linear equations is given in the order $i=n,n-1,\ldots,1$ by
\begin{equation}
2 C^{(jj)} \beta_i^{(j)} + \sum_{\underset{k\neq j}{k=0}}^{i} C^{(jk)} \beta_i^{(k)} = - C^{(ji)} \beta_i^{(i)},\quad j=0,\cdots,i-1
\label{eq:psms:linsy}
\end{equation}
with the co-variance $C^{jk} = \langle \Delta^{(j)} \Delta^{(k)} \rangle - \langle \Delta^{(j)} \rangle \langle \Delta^{(k)} \rangle$
 of the difference $\Delta^{(j)} = S^{(j)}(U')-S^{(j)}(U)$.
Implying the constrain $\beta_i^{(i)} = \beta^{(i)} - \sum_{j=i+1}^n \beta_j^{(i)}$
the linear equation system eq.~\eqref{eq:psms:linsy} can be solved, resulting into
$\beta_i^{(0)},\ldots,\beta_i^{i-1}$.
We present in table~\ref{tab:params} the optimal parameters for the runs with $d=16$ and $L=128$ at the three different gauge couplings $\beta = 3.0 , 6.0, 8.45$ with Wilson mass parameters $m_0 = -0.082626, -0.034249, 0.0$, respectively.
The parameters nicely illustrate the correlations between the different parts of the fermion action,
i.e.~roughly the full contribution of $\textrm{log} \, P^{(2)}_{j}$ is compensated within $\textrm{log} \, P^{(3)}_{j}$
and similar for $\sum_j \textrm{log}  \, P^{(3)}_{j}$ within $\textrm{log}  \,  P^{(4)}$. In general, by including fermions the gauge coupling of the pure gauge proposal shifts towards larger values. That is expected, i.e.~first contribution of the so-called hopping parameter expansion of the determinant of $D$ comes with a positive contribution of $S_g$.

\begin{figure}
  \includegraphics[width=1\linewidth]{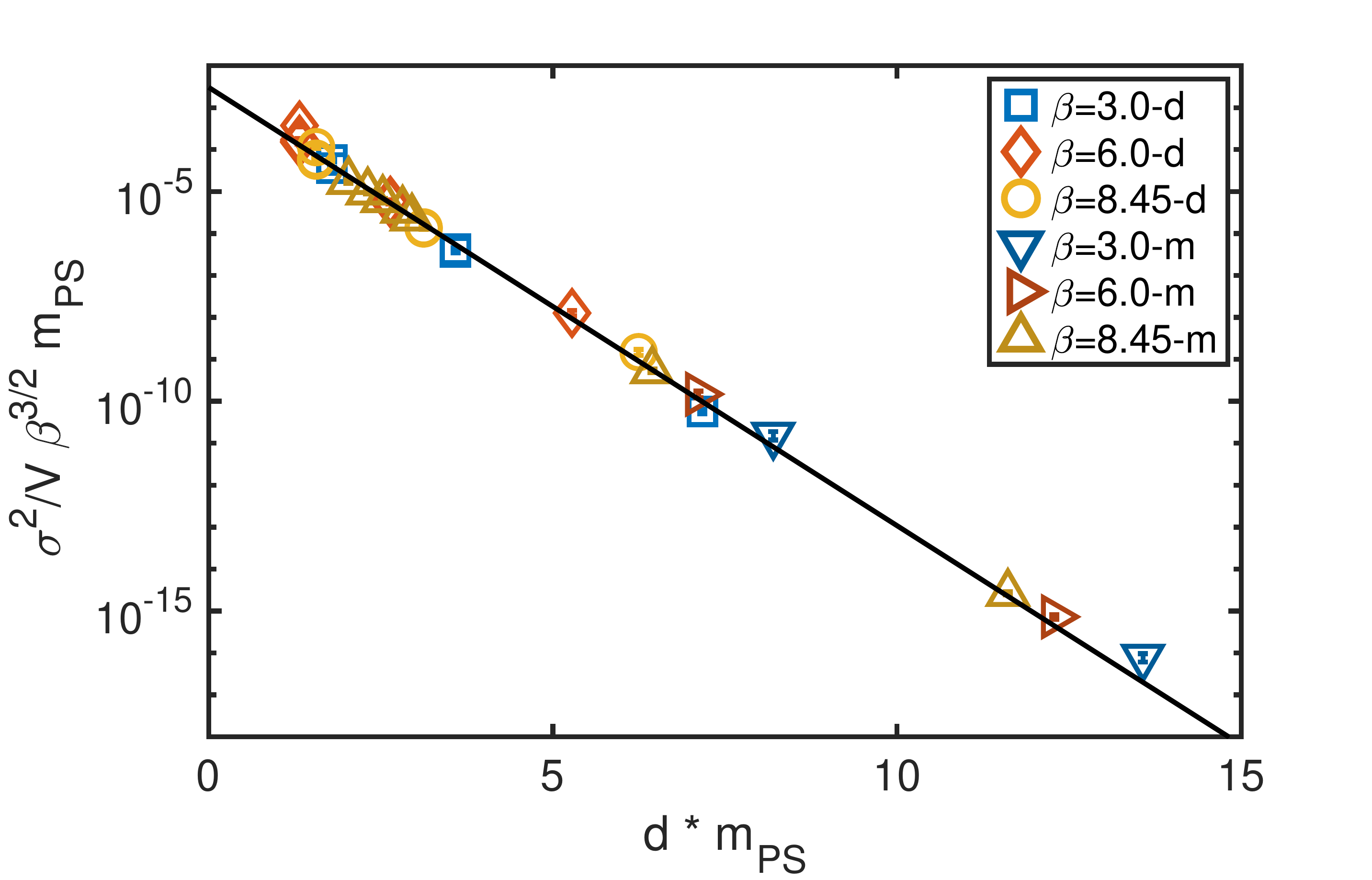}
      \caption{The fit shows the dependence of the variance of the global correction steps, within the 5 level flowGC algorithm on the distance between active blocks $d/\sqrt{\beta}$, on the pseudoscalar meson mass $m_{PS}\sqrt{\beta}$ and the volume $V/\beta$. The runs with different distances, denoted as $\textit{-d}$ are performed at parameters listed in Tab.~\ref{tab:paras} while runs with different pseudoscalar masses are denoted by \textit{-m} and performed at constant $d=16$ and lattice extend $L=64$ at the three different $\beta$ values.}
      \label{fig:sigvsmanV}
\end{figure}

\textit{Dependencies of the variance. --}
To estimate dependencies of the variance in the final global correction step of the 5 level flowGC
we generated a set of 35 simulations at different gauge couplings $\beta = 3.0, 6.0, 8.45$,
different volumes with $L=64,96,128$, different distances $d=8,16,32$
and various pseudoscalar masses between $a m_{PS} \in [0.12,\; 0.73]$ using $l=8$ gauge flow updates.
We found 
\begin{equation}
    \sigma^2 = A \cdot \frac{V}{ m_{PS} \beta^{3/2}} \; \cdot \textrm{exp}\{-B \cdot d \, m_{PS}\} 
\end{equation}
with $A=0.0030(1)$ and $B=2.4062(59)$. The dependence is depict in Fig.~\ref{fig:sigvsmanV} for the distance $d$ and the pseudoscalar mass $m_{PS}$,  while the dependence on the volume $V$ and gauge coupling $\beta$ is shown in Fig.~\ref{fig:acceptance}.
The fit in $d\cdot m_{PS}$ with $\chi^2/dof = 21$ captures the overall dependence quite well.
Note that, although we employ flow updates of domain with $l=8$, for smaller pseudoscalar
masses as well as for smaller $\beta$ the acceptance rate of the lower filter might drop
such that the overall updated fraction of the gauge links can drop under 15.6\%. This leads to a smaller variance $\sigma^2$, which is the case for most of the $\beta=3.0$ runs. This makes it 
difficult to estimate the direct pseudoscalar mass dependence, i.e.~if we optimize the leading $a m_{PS}$ dependence of $\sigma^2$ via the minimal $\chi^2$, we find $\sigma^2 \propto 1/(m_{PS} \sqrt{\beta})^x$ with $x=1.2$.
 
Note that the exponential suppression with the distance of the global filter step, lead to an \textit{exponential} increase of the fourth filter step. However, because the corresponding term only contains contribution from a single active domain, this can be mitigated via additional iterations on the last filter level.

\begin{table}[]
   \centering
    \begin{tabular}{c|c|c|c}
        $\beta$    & 3.0 & 6.0 & 8.45  \\\hline \multicolumn{4}{c}{$\phantom{.}$}   \\ 
        \multicolumn{4}{c}{5 level flowGC with $d=16$:} \\ \hline
        Level 4 \\ \hline
        with $\sigma^2$ & 0.0052 &  0.0369 & 0.0046 \\
        and $P_{acc}$  &  0.9713 &  0.9235 & 0.9727 \\ \hline
        $\delta \beta_4^{(3)}$ & -2.0037 & -2.0182  & -2.0087 \\
        $\delta \beta_4^{(2)}$ &  1.0027 & 1.0061  &  1.0083 \\
        $\delta \beta_4^{(1)}$ & -0.0003 &  0.0008 &  0.0004 \\ \\
        Level 3 & \multicolumn{1}{c}{ $n_1 = 2$}\\ \hline
        with $\sigma^2$ & 0.6688 & 0.6190 & 0.1546 \\
        and $P_{acc}$  &  0.6826 & 0.6940 & 0.8441 \\ \hline
        $\delta \beta_3^{(2)}$ & -1.1730 & -1.3635 & -1.3534 \\
        $\delta \beta_3^{(1)}$ & -0.0006 & 0.0149 & 0.0125 \\ \\
        Level 2 & \multicolumn{1}{c}{ $n_2 = 4$}\\ \hline
        with $\sigma^2$ & 1.4384 & 0.8325 & 0.1857  \\
        and $P_{acc}$  &  0.5487 & 0.6482 &  0.8294 \\ \hline
        $\delta \beta_2^{(1)}$ & -0.2482 & -0.3082 & -0.2863 \\ \\
        Level 1 & \multicolumn{1}{c}{$n_1 =100$}\\ \hline
        with $P_{acc}$ & 0.5669 & 0.2501 & 0.2794 \\\hline \multicolumn{4}{c}{$\phantom{.}$} \\
        \multicolumn{4}{c}{2 level GC:} \\ \hline
        with $\sigma^2$ & 12.3774 & 9.7119 & 3.7260 \\
    and $P_{acc}$ & 0.0786 &  0.1192 & 0.3345 \\ \hline        
    \end{tabular}
    \caption{The table shows the parameter which minimize the variance of each filter step.
    The parameter used for the runs with coupling constant $\beta=3.0,\, 6.0, \, 8.45 $ using a lattice extend of $L=128$ with a distance $d=16$ between active domains, by keeping the domain size at a similar size $L_b=16$. For comparison numbers of a two level global correction step are included, which is driven by an HMC update and a global correction step without hierarchical filter steps. \label{tab:params}}
\end{table}

\textit{Remarks on cost scaling. --}
The computational cost of the flowGC algorithm during the Monte Carlo sampling
depends mainly on the acceptance rate and the calculation of the determinant ratios.
Towards larger lattices, higher dimensions and gauge theories of larger gauge groups,
such as lattice QCD, the calculation of the determinants ratios 
requires alternative methods. 
A possible way out is based on stochastic estimation
which can be used within the hierarchical filter steps, as discussed in Ref.~\cite{Finkenrath:2012az}.
The stochastic noise scales, similar to the exact weight, with the volume.

Thus the computational cost for a gauge theory in $D$ dimensions is given by
\begin{equation}
\begin{split}
    \textrm{cost} \propto &  \left(c_{3d}  \frac{N_b \cdot n_{3} \cdot V_{3 d}}{a^D} + c_{gl} \frac{V_{gl}} {a^D}\right) \, \cdot \\ & \quad 1/\textrm{erfc}\left\{ \left[ \frac{ A'  \cdot V}{8 m_{PS} a^{D+1} } e^{-B' \cdot d \, m_{PS}}\right]^{1/2} \right\}
\end{split}
\end{equation}
with $N_b$ the number of active domains, $n_3$ the number of update iteration of filter step at level 3, with $V_{3d}/a^D$ the volume of the extended block operator, which filter out the long range fluctuations of the active blocks and $V_{gl}/a^D$ the global volume. 
The coefficients $A'$ and $B'$ are model dependent and the factors $c_{3d}$ and $c_{gl}$ are relative cost factors, e.g.~includes the total number of inversions needed for the stochastic determinant estimation.  For high acceptance the computational cost is likely to be dominated by the extended block operator with $V_{3d}/a^D$
if large distances $d$ are neccessary to compensate small pseudoscalar masses. Note that it is possible to introduce additional filter levels via recursive Schur decomposition,
if the corresponding acceptance rate drops.

\begin{figure}
  \includegraphics[width=.99\linewidth]{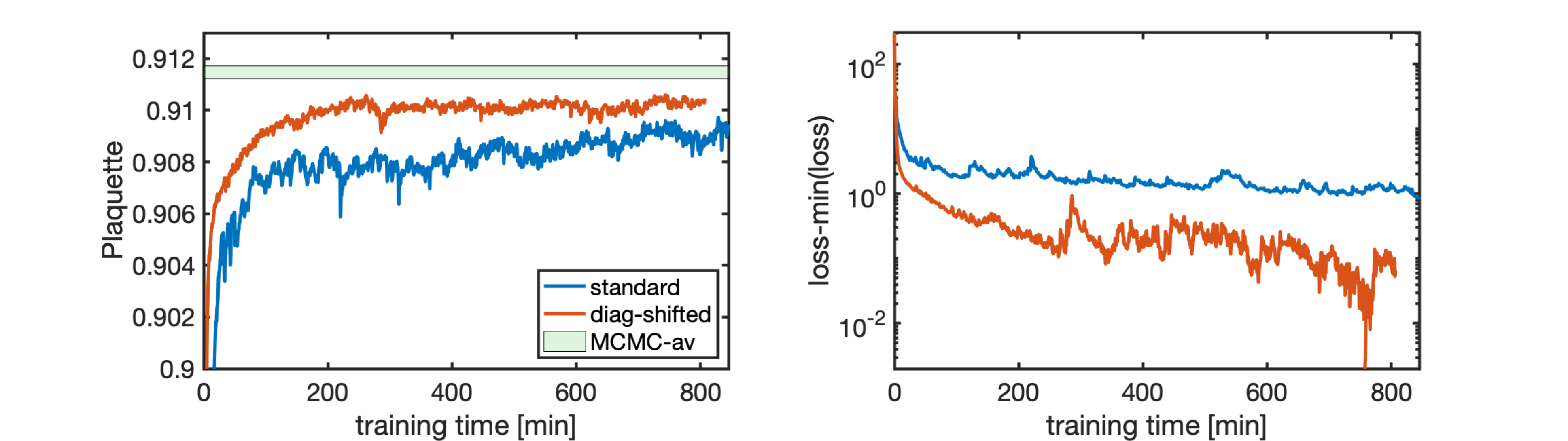}
      \caption{\label{fig:com_mask}The figure shows the comparison of the training effectiveness using the standard masks compared to the optimised version, with diagonal shifted masks. Left panel shows the plaquette average of each batch after one epoch, while the right panel depict the minimization of the loss function in dependence of the training time in case of periodic boundary condition.}
\end{figure}
\begin{figure}
\includegraphics[width=.48\linewidth]{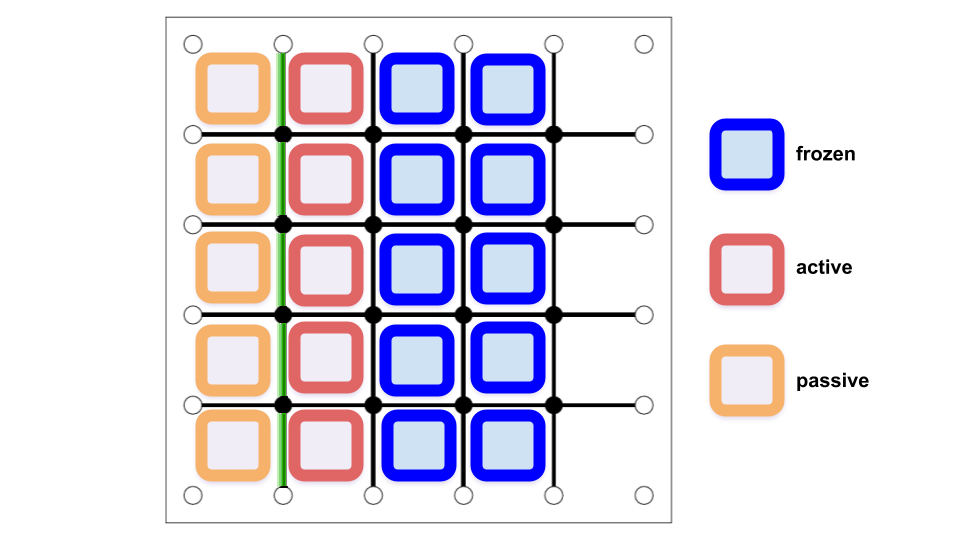}
\includegraphics[width=.48\linewidth]{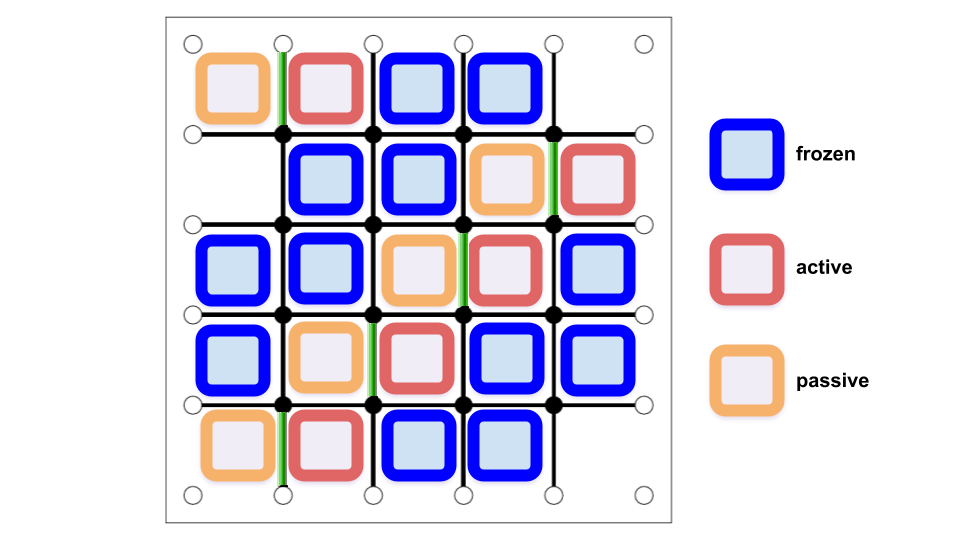}
      \caption{ \label{fig:masks2}The figure shows the mask setup used in the gauge equivariant flows. The original approach\protect{\cite{Albergo:2021vyo}} is shown in the left panel, while faster convergence can be achieved with the diagonal shifted approach (right panel). }
\end{figure}

\textit{Remarks on masks in gauge invariant flows. --}
The flow maps are constructed of coupling layers, which maps the links 
towards the target distribution. For simplified computation of the Jacobian, only a couple of links are updated in each transformation. Here, we use plaquettes as the gauge invariant objects \cite{Kanwar:2020xzo,Albergo:2021vyo,Boyda:2020hsi}. 
Using gauge invariant objects results then into three different subsets or masks within the coupling layers, namely
\begin{itemize}
    \item frozen, which are used for updating
   \item active, which are getting updated
   \item passive, which are updated passively.
\end{itemize}
Within the original proposal \cite{Albergo:2021vyo}, the lattice were divided vertical into columns,
starting with a passive column, followed by an active and two frozen columns,
as illustrated within the left panel of Fig.~\ref{fig:masks2}.
Now, the links laying between the passive and active columns can be updated based on the trainable coupling layers
with input from the frozen links.

After an update step, the maps are rotated
and the links in opposite direction are updated.
This is iterated until all links are updated,
which need an application of 8 coupling layers in two dimensions.

We increased convergence by shifting the masks diagonal,
as illustrated on right panel in Fig.~\ref{fig:masks2}.
This increases overlap with frozen plaquettes in case of
a kernel size of 3 for the employed convolutional networks within
the coupling layers and results into faster convergence at similar
training time, as shown in Fig.~\ref{fig:com_mask}.
For the training, we increased the number of layers to 64
while keeping other parameter, such as the hidden layers $(8,8)$ or the convolutional kernels size
$3$ similar to the original default values, see Ref.~\cite{Albergo:2021vyo}.


\end{document}